\documentclass[pre,aps,twocolumn,
superscriptaddress,floats,showpacs,10pt]{revtex4-1}
\usepackage{graphicx}
\usepackage{dcolumn}
\usepackage{bm}
\usepackage{amssymb}
\usepackage{amsmath}
\usepackage{textcomp}
\usepackage{color}

\newcommand{\re}[1]{{\color{black}#1}}

\begin{document}

\title{On annihilation of the relativistic electron vortex pair in collisionless plasmas}

\author{K.V. Lezhnin}
\affiliation{Department of Astrophysical Sciences, Princeton University, Princeton, New Jersey 08544, USA}
\affiliation{National Research Nuclear University MEPhI, Kashirskoe sh. 31, 115409, Moscow, Russia}

\author{F.F. Kamenets}
\affiliation{Moscow Institute of Physics and Technology, 
Institutskiy per. 9, Dolgoprudny, Moscow Region 141700, Russia}

\author{T.Zh. Esirkepov}
\affiliation{National Institutes for Quantum and Radiological Sciences and Technology, 8-1-7 Umemidai, Kizugawa, Kyoto 619-0215, Japan}

\author{S.V. Bulanov}
\affiliation{National Institutes for Quantum and Radiological Sciences and Technology, 8-1-7 Umemidai, Kizugawa, Kyoto 619-0215, Japan}
\affiliation{Institute of Physics of the Czech Academy of Sciences v.v.i. (FZU), Na Slovance 1999/2, 18221, Prague, Czech Republic}
\affiliation{A. M. Prokhorov Institute
of General Physics of the Russian Academy of Sciences, Vavilov Street 38, Moscow, 119991, Russia.}

\date{\today}

\begin{abstract}
In contrast to hydrodynamic vortices, vortices in plasma 
contain an electric current circulating around the center of the vortex, which generates a magnetic field localized inside. 
Using computer simulations, we demonstrate that the magnetic field associated with the vortex gives rise to a mechanism 
of dissipation of the vortex pair in a collisionless plasma, leading to fast annihilation of the magnetic field  
with its energy transforming 
into the energy of fast electrons, secondary vortices, and plasma waves. \re{Two major contributors to the energy damping of double vortex system, namely, magnetic field annihilation and secondary vortex formation, are regulated by the size of the vortex with respect to the electron skin depth, which scales with the electron gamma-factor, $\gamma_e$, as $R/d_e \propto \gamma_e^{1/2}$. Magnetic field annihilation appears to be dominant in mildly relativistic vortices, while for the ultrarelativistic case, secondary vortex formation is the main channel for damping of the initial double vortex system.} 

\bigskip

\end{abstract}

\maketitle

\section{Introduction}

Formation and evolution of localized nonlinear structures such as vortices and solitons play a crucial role in the physics of continuous media \cite{VORTEXBOOK, SOLITONBOOK}. 
For instance, drift wave dynamics in tokamak plasmas can be described within the framework of the Hasegawa-Mima (HM) equation \cite{HMEQN}, which has a well-known point vortex solution. The vortices may affect energy and particle transport significantly \cite{KONO, KRASH}. The formation of finite-radius relativistic electron vortex structures 
associated with quasistatic magnetic field generation provides one of the pathways for the electromagnetic field energy  depletion in laser plasmas \cite{VORTEX}. 
The late stage of the vortex evolution resulting in strong plasma density modulations has been revealed in the experiments  \cite{VORTEXEXP} using proton radiography. Electron vortex pairs are also observed in simulations of relativistic shocks, being responsible for electron energization in the upstream region \cite{Naseri2018}. Understanding the dynamics of vortex structures \re{in plasmas} is important for developing the theory of relativistic plasma turbulence \cite{TURBPLASMA2}. \re{Relativistic electron vortex dynamics may also be a significant factor in the late stages of relativistic Weibel-like instability, which can arise in superstrong laser-plasma interaction \cite{Wei2004}, as well as in colliding astrophysical flows of electron-positron plasmas \cite{Kazimura1998}.}

In contrast to hydrodynamical vortices, which are sustained by  fluids {comprised of neutral particles}, vortices in plasmas are sustained by the rotational motion of charged particles, leading to nonzero circular electric current, which forms a magnetic field inside the vortex \cite{LEZHNIN}. 
In the case of small radius vortices, which correspond to the  point-vortex solution of the HM equation, the vortex internal energy is conserved during the interaction process.
However, in the case of finite radius vortices, we expect the finite-radius and electromagnetic interaction effects to become prominent, 
leading to a fast vortex energy  dissipation with its transformation into the energy of fast particles. Below, using two dimensional (2D) Particle-In-Cell (PIC) simulations with the code REMP \cite{REMP}, 
we demonstrate how pairs of vortices interact beyond the point vortex approximation. 
We reveal the effect of relativistic annihilation of the binary electron vortices magnetic field that leads to vortex pair dampening.

\section{Simulation setup}

The simulation parameters are as follows. For clarity, we describe the simulation setup in terms of an arbitrary spatial scale parameter, $\lambda$, and then immediately rescale the model to the physically relevant units. We set a slab of electron plasma (assuming immobile ions) with a constant density gradient along the $x$ axis, 
so the electron plasma density equals $n_e/n_{\rm max} = 0.1$ at $x = 55 \lambda$ and $n_e/n_{\rm max} = 1$ at $x = 95 \lambda$, with width $40 \lambda$ and zero temperatures for electrons. 
We measure spatial parameters in $\lambda$, temporal -- in $2 \pi / \omega_0 = \lambda / c$, densities -- in $n_{\rm 0}=m_e \omega_0^2 / 4 \pi e^2$, 
electromagnetic fields -- in $E_0 = m_e \omega_0 c / e$, where $m_e$ is electron mass, $e$ is the absolute value of electron charge, $c$ is the speed of light in vacuum. {For the sake of simplicity, we introduce circularly symmetric electron vortices. They are initiated by accumulating the localized magnetic field during a number 
of timesteps at the beginning of the simulation \cite{SSBULANOV, LEZHNIN}. For the simulations presented, electron vortices are formed with various maximum magnetic fields: $B_{\rm max} = 0.5, 1, 2, 4, 6.5, 35$ in plasma with $n_{\rm max}=0.16, 0.36, 0.64, 1, 4, 16$,
 respectively. Hereafter, we will refer to the simulation parameters by the magnetic field amplitude $B_{\rm max}$. The vortex centres are located around points $x=75 \lambda$ and $y= -4 \lambda, 4 \lambda$. We choose our parameters in such a way that the condition $\omega_{pe}^2 \ll \omega_B ^2$ holds \cite{Gordeev}, 
so the electrons can be considered magnetized. Here $\omega_{\rm pe}^2=4 \pi n_e e^2/m_e$ is the plasma frequency and $\omega_B = e B / m_e c$ is the electron gyrofrequency. The computational grid is $150 \lambda \times 120 \lambda$ with 32 nodes per $\lambda$, boundary conditions are periodic. We have also qualitively verified the results of our simulations with a larger domain resolution (64 and 128 nodes per $\lambda$). The initial particle-in-cell number corresponding to the maximum electron density is equal to 100. The total number of particles is about $10^8$.
The integration timestep is 0.0155. The total time of the simulations is 500 time units. 

\re{For the sake of clarity, we further rescale our numerical model to physically relevant units appearing from the simple electron vortex model. It can be formulated as follows. Let us assume that the electron moves in a circular orbit around the uniformly distributed immobile and positively charged ions. Then, the electric field experienced by the electron is $E=2 \pi e n R$, where $R$ is the radius of the electron vortex and $n$ is the ion density. Assuming the electron to have a speed $v_e \approx c$, we obtain the magnetic field to be $B=2 \pi e n R$. Radial force balance for the electron can be written as $v_e p_e /R = -e E$, which gives an expression connecting electron vortex radius and electron momentum, $R = (p_e c / 2 \pi n e^2)^{1/2} \approx d_e \sqrt{2 \gamma_e}$. Thus, we fix $\lambda = (4\pi^2  n_{\rm max}/n_{\rm 0} \cdot m_e c /p_e)^{1/2} R$, normalizing all spatial quantities to $R$, temporal frequencies to crossing frequency $\omega_{cr} = c/R$, fields to $E_0' = m_e \omega_{cr} c/ e$, densities -- to $n_{\rm 0}'=m_e \omega_{cr}^2 / 4 \pi e^2$.}

\section{PIC simulation results and theoretical estimates}

In our simulations, we expect {to observe} the following scenario: first, when two vortices are far away from each other ($> 5 R$), 
they would be stationary unless we were to take into account the effects of a finite vortex radius. 
In the latter case, we can expect that the vortices will move perpendicularly to the density gradient (parallel to the y-axis), 
due to the conservation of the Ertel's invariant $I = \Omega / n$, where $\Omega$ is the vorticity and $n$ is the electron density \cite{ERTEL}. 
The velocity of such motion is estimated as $\Omega R^2 |\nabla n/n|$, which is $\lessapprox c /80$ and has turned out to be fairly consistent with the simulation results presented below. Then, when the vortex interaction becomes significant (it scales as $K_0(|\Delta y/d_e|)$ with the vortex separation $\Delta y$, $K_0$ is the modified Bessel function of second kind, see, e.g., \cite{KONO}), we expect the binary vortex to start moving 
along the $x$ axis and possibly follow one of the complicated trajectories discussed in Ref. \cite{KONO}. The typical velocities of such motion are $V_{\rm bin} \approx 0.2 - 0.5 c$.
Eventually, the vortex binary tightening until $\sim R$ 
will lead to the finite-radius effects coming into play, which are beyond the scope of applicability of the point vortex theory described in Ref. \cite{KONO}. 
To reveal the finite vortex radius effects and the effects of magnetic interaction we perform the PIC simulations.

\begin{figure}
    \includegraphics[width=0.99\linewidth]{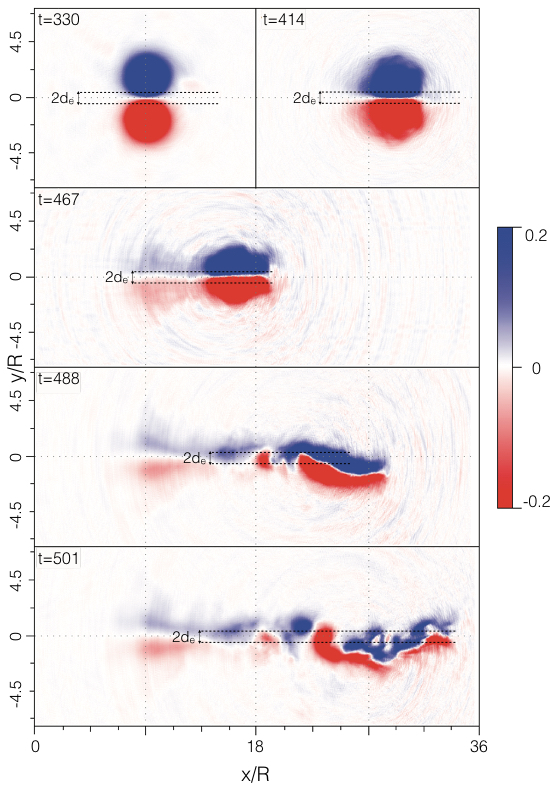}
\caption{Sketch of the binary vortex evolution - z component of the magnetic field: approaching each other (t=330), formation of the dipole vortex structure (t=414), 
radiation of electromagnetic waves and formation of a dipole magnetic field structure in the wake of the dipole vortex (t=467), 
decay of the dipole vortex into smaller electron vortices, which form von Karman vortex rows (t=488), and the magnetic field annihilation, leading to electron heating (t=501). $2 d_e$ width scale, tightly connected to the annihilation process, is demonstrated.}
\label{fig:init}
\end{figure}

Figure 1 {illustrates typical } evolution of the $B_z$ component of the magnetic field {observed} during the simulation (for $B_{\rm max}=2$). When the binary vortex system is tight enough (i.e. distance between the closest points of the vortices is $\sim d_e$, where $d_e=c/\omega_{\rm pe}$ is the electron skin depth, Fig. 1, t=330), the point vortex approximation breaks down. The electron currents of the two vortices, both directed along the $x$ axis in the closest point of approach, attract each other and form a magnetic-dipole vortex structure (Fig. 1, t=414) \cite{NAKAMURA}. The structure observed has an analogue in hydrodynamics, which is known as the Larichev-Reznik dipole vortex solution \cite{L-D}. This type of structure is believed to be stable in the hydrodynamic case \cite{HYDRVORT}. However, in our case, the magnetic structure moves along the $+x$ direction, losing the majority of its magnetic energy by turning it into electromagnetic waves (Fig. 1, t=467; Fig. 3b), accelerated electrons and forming of von Karman-like streets of secondary vortices (Fig. 1, t=488, 501; Fig. 3b, Fig. 3c), though, secondary vortex formation does not decrease the total magnetic energy of the system significantly. The direction of the binary vortex motion may be deflected from the straight propagation along the $x$ axis, as the binary components disintegrate unequally on the secondary vortices, and the resulting binary vortex with unequal components deflects in the direction of the larger vortex component, in agreement with \cite{KONO}. The rapidly accelerated electrons are a sign of the relativistic magnetic field annihilation. The annihilation of the magnetic field was observed in PIC simulations previously in a different geometry \cite{YGU} between the azimuthal magnetic fields formed by two parallel laser pulses propagating in a nonuniform underdense plasma and leads to electron heating. Though the overall physics of the Ampere's law is the same in both cases, as well as the signature of rapid electron energization, in \cite{YGU} the displacement current arose as a result of the magnetic fields expanding towards each other due to the negative density gradient along the propagation axis of the laser pulses. In our case, the two vortices are pushed towards each other by the finite-radius effect of the vortex drift motion. Still, in both cases the dynamics of the magnetic fields is guided by the conservation of the Ertel's invariant. The process of secondary vortex formation may be caused by vortex boundary bending, observed in simulations previously \cite{LEZHNIN}. Secondary vortices are not subject to the vortex film instability \cite{VORTEX}, as the finite vortex radius effects dominate the motion of the vortices which are separated by a few $d_e$. The role of the relativistic effects is demonstrated using auxiliary simulations with $n_{\rm max} = 0.36$ with a large range of $B_{\rm max}$ from 0.1 to 2. It was demonstrated that the magnetic field damping in the nonrelativistic case is at least three times longer, and the electric fields coming from the displacement current term in Ampere's law are negligible, see \cite{Wang2017}. 

A simple model of the magnetic field annihilation of electron vortices may be written as follows. The radius of a vortex is connected to the electron momentum by relation $R/d_e = (2 p_e / m_e c)^{1/2}$. Thus, the nonrelativistic vortices have radius $R \leq d_e$ and the ultrarelativistic vortices have $R \gg d_e$.  Ampere's law is generally stated as $\nabla \times \mathbf{B} = 4\pi/c \cdot \mathbf{J} + 1/c \cdot \partial \mathbf{E}/\partial t$. It may be rewritten as an order-of-magnitude estimate, using $|\nabla \times \mathbf{B}|\approx |\partial B / \partial y| \sim |B / d|$, where $d$ is the typical spatial gradient scale length,  $|\mathbf{J}|\approx e n_e c$ for the limit when $v_e \sim c$,  $|\partial \mathbf{E} /\partial t| \sim E / \tau$, where $\tau$ is the typical temporal scale. Finally, it yields $d/d_e = B / (1 + E/ \omega_{\rm pe} \tau)$ (B and E are dimensionless). Thus, it is clear from this equation that reaching $d_e$ scale ($d \leq d_e$) is necessary for the magnetic field annihilation through the displacement current term (see, e.g., \cite{YGU,Wang2017}). Thus, the more relativistic the vortex is (in terms of $p_e/m_e c \approx \gamma$ parameter), the harder it is to squeeze the dipole vortex down to a $d_e$ scale. That being said, large vortices (in terms of $d_e$ scale) are harder to damp via the magnetic field annihilation.

Let us compare two types of simulations with the same parameters except for the signs of the magnetic fields in the vortices. Thus, in one case the vortices move towards each other and interact (Figure 2, blue line), in the other case they move away from each other and do not decay on the timescale of the simulations (Figure 2, dashed black line). Figure 2 shows the rate of magnetic energy dissipation in both simulations. Here, we can distinguish at least two mechanisms of vortex dissipation - slow (dashed lines, dissipation time is larger than $10^3 \omega_{cr}^{-1}$) and fast (solid lines, typically less or much less than $10^3 \omega_{cr}^{-1}$). The first mechanism can probably be attributed to the formation of spiral density waves in the electron plasma, which are seen in the early stages of simulations  (e.g., see spiral perturbations of electron density in Fig. 3a and Fig. 4a). In our simulations, this mechanism gives us the rate of dissipation which dissipates no more than 20\% of the magnetic energy during the simulation time, so it will not impact the characteristic lifetime of the electron vortex, or at least will make a contribution on a longer timescale than the fast dissipation, which will be discussed below. In turn, fast vortex dissipation can destroy the vortex pair on a much shorter timescale. Synchrotron losses, in comparison to electromagnetic solitons, are also negligible in the electron vortex case \cite{Esirkepov2004}.

As the result of the magnetic energy dissipation, we observe a bunch of electrons being accelerated approximately in $+x$ direction, adding up to $\sim 60 m_e c$ to the electron momentum in comparison to the maximum electron momentum of the stationary electron vortices in the case of $B_{max} = 35$. Figure 4 demonstrates the effect of the electron acceleration. The energy of electrons is large enough for the bunch to escape the plasma region. According to Figure 2, we see that the more relativistic vortices, with larger $\gamma$-factors, are harder to annihilate, in agreement with our theoretical model. Secondary vortices, which are more prominent in the simulations with higher $\gamma$ factors of the initial vortices, are also more stable against the magnetic field annihilation, which results in the saturation of the magnetic field energy in the system (see Figure 2, aqua and purple lines).

It is also important to note that the immobile ion approach is justified only if $\omega_{\rm pi}/\omega_{\rm pe} \ll 1$ and $2 \pi / \omega_{\rm pi}$ is greater than the total simulation time. Besides, the binary vortex motion should be fast enough so we could ignore the ion motion: $V_{\rm bin}/R \gg \omega_{\rm pi}$, where $R$ is the typical radius of the vortex. Otherwise, the binary system of vortices does not move according to the HM equation, but they evolve independently \cite{LEZHNIN} until two vortex boundaries collide. The effects of ion inertia on the binary vortex system will be considered in a separate paper.

\begin{figure}
    \includegraphics[width=0.99\linewidth]{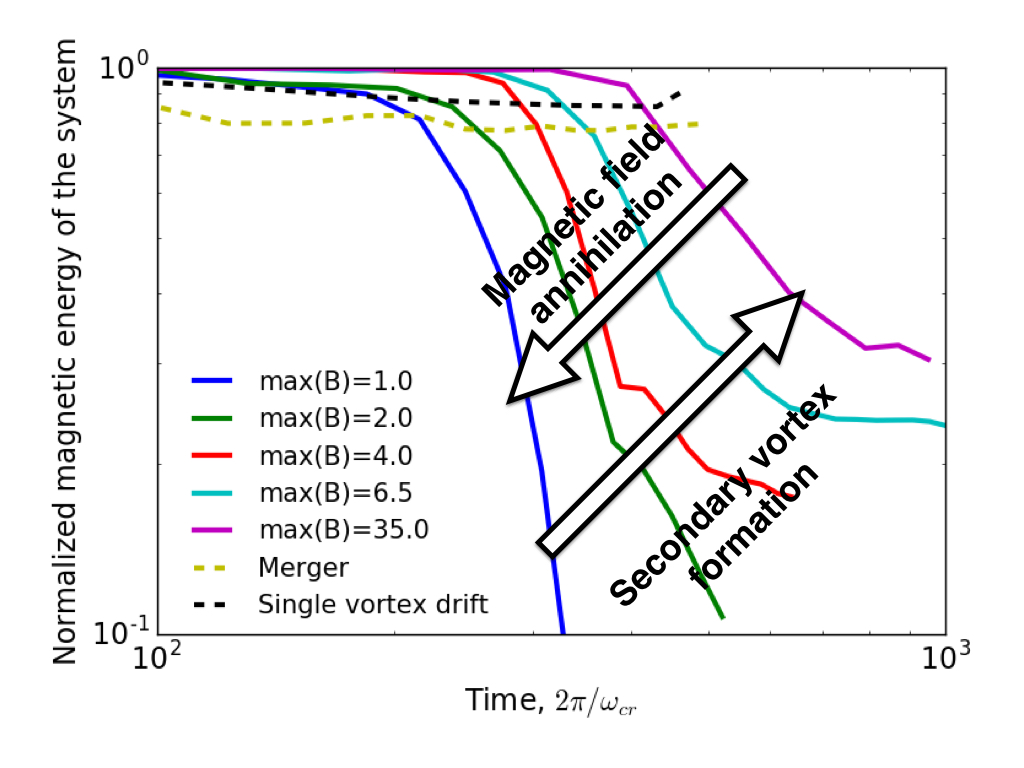}
\caption{Normalized magnetic field energy evolution over time for various cases - drifting single vortices (dashed black line), merging vortices (dashed brown line), dissipating vortices for $B_{max}=1.0$ (blue), $B_{max}=2.0$ (green), $B_{max}=4.0$ (red), $B_{max}=6.5$ (aqua), and $B_{max}=35$ (purple). In the case of smaller vortices (in terms of $R/d_e$) the magnetic field annihilation dominates the vortex damping, in the case of larger density values - secondary vortex formation mitigates the total magnetic energy dissipation.}
\label{fig:emagn}
\end{figure}

\begin{figure}
    a)\includegraphics[width=0.97\linewidth]{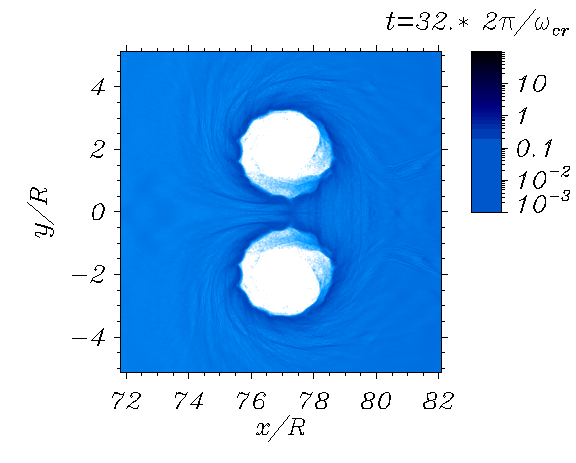}\par
    b)\includegraphics[width=0.97\linewidth]{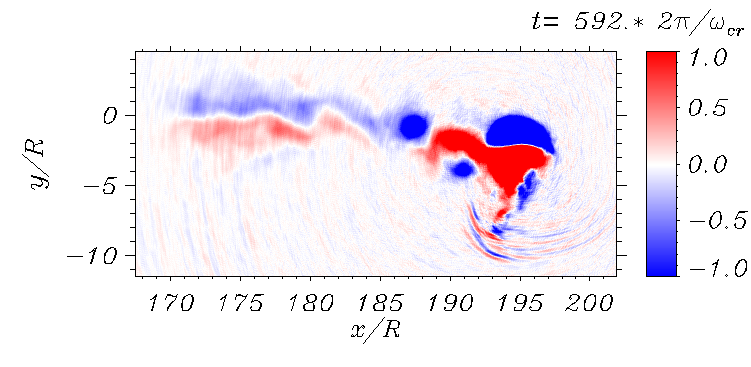}\par
    c)\includegraphics[width=0.97\linewidth]{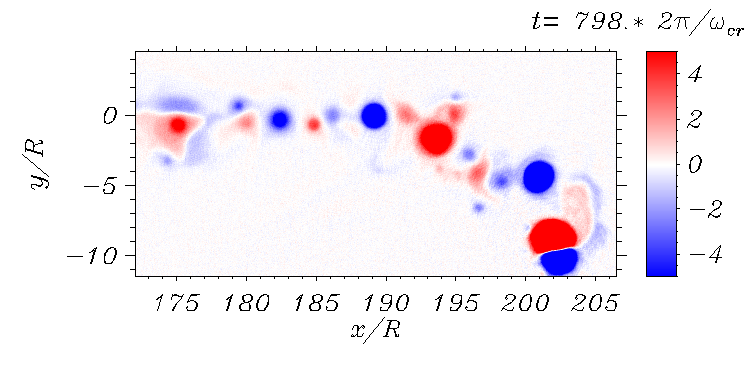}\par
\caption{a)  Electron density distribution for t=32 (simulation with $B_{max} = 0.5$). Spiral density waves, which possibly correspond to the electromagnetic energy dissipation mechanism on early stages of vortex evolution, are seen; b) $B_z$ component of the magnetic field for t=592 (simulation with $B_{max} = 6.5$). Around $x=193$ and $y=-8$ we observe the emission of the electromagnetic wave. c) $B_z$ component of the magnetic field for t=798 (simulation with $B_{max} = 35.0$). The von Karman-like street of secondary vortices is observed in the wake region of the dipole vortex.}
\end{figure}

\begin{figure}
    a)\includegraphics[width=0.97\linewidth]{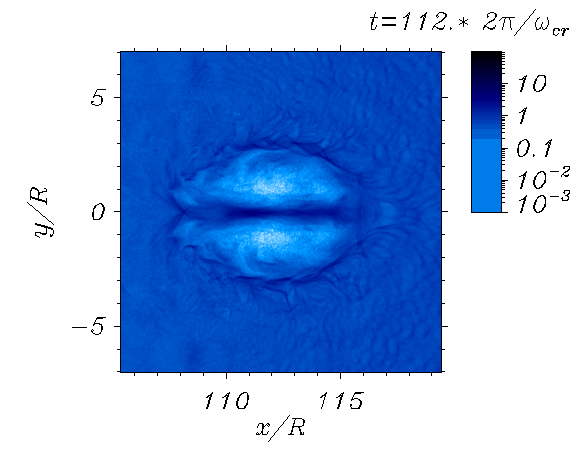}\par
    b)\includegraphics[width=0.97\linewidth]{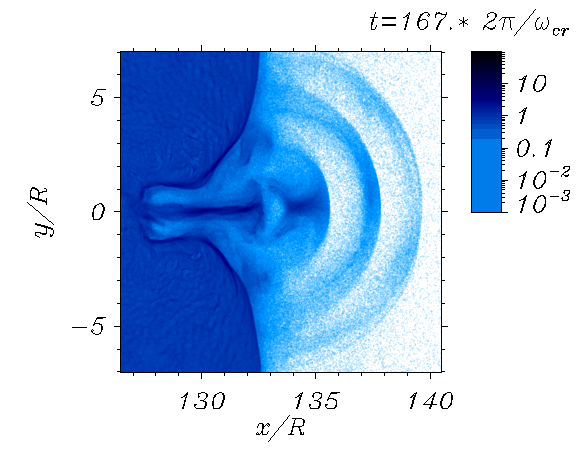}
\caption{Electron density distribution at a)t=112 and b)t=167 ($B_{max}$=4 simulation). The annihilation of the magnetic field leads to the formation of an electron bunch with an energy allowing to escape the plasma into vacuum.}
\end{figure}



The simulation setup used in the problem, such as a plasma density gradient, is implemented in order to consider the adiabatic switching-on of the vortex interaction effects. Thus, we may observe the same effect of vortex damping in homogeneous plasmas when forming tight binary systems of vortices using our numerical scheme. However, in order to exclude the effect of the initial generation process, which inevitably will cause strong coupling between the vortex pair, and demonstrate the stability of single electron vortices, we decided to form vortices far away from each other, making sure that the vortex generation process does not impact their interaction and the magnetic field energy is almost constant over the simulation time (for non-interacting vortices). The dashed black line
in Figure 2 demonstrates the evolution of the magnetic energy in the single vortex drift case. In general, the lifetime of the electron vortex binaries in the homogeneous plasma appears to be longer than in the nonzero density gradient case. 

It is also natural to discuss a system of binary vortices with the same polarization of magnetic field. In the point vortex approximation, they will simply rotate around each other in the case of homogeneous plasma \cite{KONO}. However, it turns out that the finite radius vortices are subject to a merger process, 
which may also lead to minor electromagnetic energy dissipation (Figure 2, dashed brown line) via the spiral density wave formation 
by the resulting ellipsoidal vortex \cite{LEZHNIN_SPIE}, which turned out to be in principle agreement with the results of the hydrodynamical simulations of the 2D vortex merger process \cite{MERGER}.

\section{Conclusions}

In conclusion, we presented the computer simulation results on the interaction of electron vortex binaries. 
These structures are often seen in 2D PIC simulations of various laser-plasma configurations and are crucial for understanding the superstrong magnetic field {evolution} and turbulence{ in relativistic plasmas}. 
If the binary vortex system is tight enough, the point vortex approximation breaks down, 
and the binary vortex is subject to the fast annihilation. 
The vortex annihilation leads to acceleration of the electron bunches, which {in its turn} leads to propagating electrostatic {waves}. 
In the case of larger $\gamma$ factor of the initial vortices (i.e., for simulations with $B_{max}=4$ and more), we also observe formation of the von Karman-streets of secondary vortices, the motion of which is stabilized by the drift motion due to the finite-radius effects. \re{Mildly relativistic electron vortex pairs damp mainly through the annihilation of the magnetic field, while ultrarelativistic electron vortex pairs decay via the secondary vortex formation.}
{We believe that the results obtained} will be useful for the development of a theory describing electromagnetic turbulence in relativistic plasmas \cite{Naseri2018,TURBPLASMA2}.

This work utilized the MIPT-60 cluster, hosted by Moscow Institute of Physics and Technology (we thank Ilya Seleznev for running it smoothly for us). 
SVB acknowledges support at the ELI-BL by the project High Field Initiative  (CZ.02.1.01/0.0/0.0/15\_003/0000449)
from European Regional Development Fund. KVL is grateful to ELI-Beamlines project for hospitality during the final stages of this work. KVL thanks Veniamin Blinov for fruitful discussions.

\end{document}